# A simple derivation of Kepler's laws without solving differential equations.

**J-P Provost**[1] **and  C Bracco**[2,3]

[1]INLN, Université de Nice-Sophia Antipolis, 1361 route des lucioles, Sophia-Antipolis, 06560 Valbonne, France

[2] Syrte, CNRS, Observatoire de Paris, 61 avenue de l'Observatoire, 75014 PARIS

[3] UMR Fizeau, Université de Nice-Sophia Antipolis, CNRS, Observatoire de la Côte d'Azur, Campus Valrose, 06108 Nice Cedex, France

E-mail: cbracco@unice.fr

**Abstract:**
Proceeding like Newton with a discrete time approach of motion and a geometrical representation of velocity and acceleration, we obtain Kepler's laws without solving differential equations. The difficult part of Newton's work, when it calls for non-trivial properties of ellipses, is avoided by the introduction of polar coordinates. Then a simple reconsideration of Newton's figure naturally leads to an explicit expression of the velocity and to the equation of the trajectory. This derivation, which can be fully apprehended by beginners at university (or even before) can be considered as a first application of mechanical concepts to a physical problem of great historical and pedagogical interest.

**1. Introduction.**

The discovery of Kepler's laws concerning the motion of celestial bodies and their explanation by Newton, have always been an integral part of scientific culture. As well known, after his analysis of Tycho Brahe's precise observations, Kepler formulated the following three propositions: -The orbit of every planet is an ellipse with the sun at a focus; -A line joining a planet and the sun sweeps out equal areas during equal intervals of time; -The square of the orbital period of a planet is directly proportional to the cube of the semi-major axis of its orbit. Newton's explanations, in the *De Motu corporum in gyrum* of 1684 [1], of these laws from the hypothesis of a central attractive force which is proportional to the inverse square of the distance to the attracting body, has been considered as one of the greatest achievements in the history of physics. However if the description of the motion of planets and the expression of the gravitational force are generally taught at secondary school, the derivation of Kepler's laws from the fundamental law $F = ma$ of mechanics only appears in first year's university courses in physics. The reason of this late introduction is that, at least in its most widespread form (see e.g. [2]), it needs a good control of differential calculus. Our main goal in the present paper is to show that this mathematical prerequisite is avoidable, which makes the deduction of Kepler's laws more accessible.

Let us recall that Newton himself did not use differential calculus in the *De Motu* (as well as in the corresponding sections of the *Principia* [3]). He replaced the continuous effect of a force by a succession of impulses, occurring at regularly distributed instants between which the motion is inertial. This approach which we shall reproduce in sections 2 and 3, allowed him to derive the law of equal areas very simply from the observation of a geometrical figure. But Newton's use of this law in order to associate the Keplerian ellipses with a $r^{-2}$ force lay on a thorough knowledge of the geometrical properties of ellipses, which is not that of present students. Further the succession of steps in Newton's reasoning is far from being intuitive (see e.g. De Gandt in [1] p. 232). In section 4, while keeping Newton's discrete time approach, and his original figure, we show that the introduction of polar coordinates brings a noticeable simplification because each sudden variation of speed $\delta v$ can be identified with that of the orthonormal unit vector in the following time interval. This makes the student rediscover by geometrical means the existence of a (not so well) known constant vector[4], which directly leads to the equation of the trajectory. In section 5, we finally show that this discrete time approach also gives the correct formula for the orbital period.

## 2. Geometrical representations of speed, acceleration and motion.

In a discrete time description of the motion of a body, let *A* be some (arbitrary) position at time $t_0$ and *B*, *C* etc. be the positions at times $t_1 = t_0 + \Delta t$, $t_2 = t_1 + \Delta t$, etc (figure 1). We suppose that the time interval $\Delta t$ is small enough such that *A*, *B*, *C*, etc. are located sufficiently close to each other to sample the trajectory. Once this interval $\Delta t$ has been taken, it is natural, in a first approach to Mechanics, to take it as the unit of time. The reason is that it allows describing the notions of speed and acceleration in very simple terms. For example, the speed between $t_0$ and $t_1$ is represented by the vector ***AB*** according to the usual sentence "speed is the (vectorial) space interval which is traveled per unit time". Let us note that this sentence is generally introduced to paraphrase the mathematical definition of the instantaneous speed (a non trivial concept with a long history [6]). Its role is clearly to soften the difficulty of the concept of derivative $v = dr/dt$ (limit $\Delta t \to 0$ of the quotient $\Delta r/\Delta t$) for students when they are first confronted to it. Here we take this sentence literally.

If the motion is an inertial one, i.e. if it proceeds at constant speed, the positions *B*, *c* etc. at times $t_1$, $t_2 = t_1 + 1$ etc. are regularly placed on the line (AB): ***AB***=***Bc***, etc. A non inertial motion implies by definition a change of speed. In physics lectures (e.g.[2]), this change is characterized by the notion of acceleration which is the limit $\Delta t \to 0$ of $\Delta v/\Delta t$ (***v*** itself having been already defined as a limit). The corresponding usual (softening) sentence is "acceleration is the speed change per unit time". If one takes also this sentence literally, its mathematical expression is

$$a = \delta v .\qquad(1)$$

Geometrically it means that if ***v*** =***AB*** is the speed between *A* and *B*, the speed between *B* and *C* is ***Bc***+***a*** (since ***AB***=***Bc***).

In this discrete time approach, it is natural to consider that the acceleration has occurred at *B*, and is obtained from the force acting at this point. Figure 1, which is based on Newton's one[5] in Theorem I [1], illustrates the construction of *C* in the case of gravitation where ***a*** points towards the force center *S* (the Sun). As Newton said: "But when the body arrives at *B*, let the central force act by a unique but strong impulse and forces the body to deviate from the line *Bc* and to continue on line *BC* ». In modern terms, Newton's construction corresponds to the algorithm[6] $r_{n+1} = r_n + v_n$, $v_{n+1} = v_n + f_{n+1}$ where $r_n$ is the position at time $t_n$, $v_n$ is the velocity between times $t_n$ and $t_{n+1}$, and $f_{n+1}$ is the force at position $r_{n+1}$. The growing development of numerical calculations in physics teaching already emphasizes the importance of figure 1. We now see how Kepler's first and second law also follow from it.

---

[4] This constant vector is a first integral of the motion, which is equivalent to the Laplace-Hamilton-Runge-Lenz vector. For a history of these integrals of motion and of their use in teaching, see [4,5] and references therein.

[5] More precisely, it differs by the addition of the dotted lines and of the notations concerning vectors and polar coordinates. We have also introduced the point *a'* in order to show that when the force does not depend on the velocity, any trajectory can be traveled in the reverse direction. Indeed, starting from *C* with a velocity ***CB*** one would reach *a'* if the motion were inertial, and *A* if the acceleration is applied at *B*, i.e. if ***a'A*** = ***cC***.

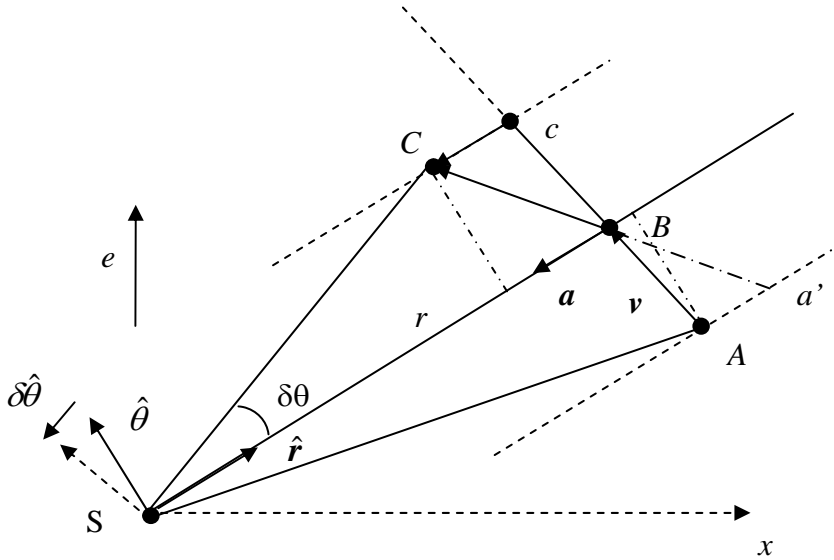

Figure 1. Newton's figure with definitions of speed, acceleration and polar coordinates.

### 3. Central force and the law of equal areas (Kepler's second law).

The hypothesis of a central acceleration immediately leads to simple well known consequences. Since *Cc* is parallel to *SB*, *C* belongs also to the plane *SAB* (plane of the figure). In the same way, since the acceleration at *C* points towards *S*, etc. the whole trajectory is a planar one. The second consequence established by Newton in Theorem I, is that the areas of the triangles *SAB* and *SBC* are the same. Indeed these triangles have a common basis *SB*, and the distances of *A* and *C* to it are equal since $AB = Bc$ and *Cc* is parallel to *SB*. The constancy of this area is Kepler's second law.

In order to obtain an analytic expression, let $r = SM$ be the distance from *S* to a point *M* (*A*, *B*, *C*, etc.) of the trajectory, and let $\delta\theta$ be the (small) angle of rotation of *SM* in the following unit of time[7] (*r* and $\delta\theta$ are represented for $M \equiv B$). The area *SBC* being approximately $\frac{1}{2}r^2\delta\theta$ (area of an angular sector) Kepler's second law for the sampled trajectory reads:

$$r^2\delta\theta = K \qquad (K = \text{Cste}). \qquad (2)$$

---

[6] See [7] for a brief history of this algorithm (which is called Newton-Hooke's algorithm by the authors) and for a comparison with some others.

[7] One could also choose for $\delta\theta$ the angle of rotation of *SM* in the preceding unit of time. What is important is to take the same convention for all positions *M*, that is, to be in conformity with the idea that figure 1 corresponds to an algorithm.

**4. The inverse square law and Keplerian trajectories (Kepler's first law).**

For the $r^{-2}$ law hypothesis, the centripetal acceleration can be written:

$$\boldsymbol{a} = -\frac{\alpha}{r^2}\hat{\boldsymbol{r}}, \qquad (3)$$

where $\hat{\boldsymbol{r}}$ is the unit vector which points from $S$ to $M$ (see figure 1) and $\alpha$ is a constant. The definition (1) of the acceleration and the law (2) of equal areas immediately lead to the relation:

$$\delta\boldsymbol{v} = -\frac{\alpha}{K}\delta\theta\hat{\boldsymbol{r}}. \qquad (4)$$

The remarkable property of this relation is that its right hand side is itself the variation of a vector. Indeed, if $\hat{\boldsymbol{\theta}}$ is the orthonormal vector deduced from $\hat{\boldsymbol{r}}$ by a $+\frac{\pi}{2}$ rotation, figure 1 shows that when $\hat{\boldsymbol{r}}$ and $\hat{\boldsymbol{\theta}}$ are rotated by $\delta\theta$ (i.e. when $B$ comes to $C$), the variation of $\hat{\boldsymbol{\theta}}$ is in the opposite direction to $\hat{\boldsymbol{r}}$ and equal in modulus to $\delta\theta$. Then formula (4) becomes:

$$\delta\boldsymbol{v} = \frac{\alpha}{K}\delta\hat{\boldsymbol{\theta}}. \qquad (5)$$

As a consequence the vector $\boldsymbol{v} - \alpha K^{-1}\hat{\boldsymbol{\theta}}$, whose variation is zero, is a constant one. This vector, which might be called Herman's vector[8], can be chosen such that the speed reads

$$\boldsymbol{v} = \frac{\alpha}{K}\left(\hat{\boldsymbol{\theta}} + \boldsymbol{e}\right) \qquad (\boldsymbol{e}\ \text{constant}). \qquad (6)$$

Relation (6) has many applications such as the construction of Hamilton's hodograph[9] or the derivation of Rutherford's diffusion angle (see for instance [8]). But the most noticeable one, in a first approach to Keplerian motions, is that it immediately leads to the equation of the trajectory. Indeed, the projection of the speed $\boldsymbol{v} = \boldsymbol{AB}$ (or $\boldsymbol{v} = \boldsymbol{CB}$) on $\hat{\boldsymbol{\theta}}$ is $r\delta\theta$ (from figure 1) $=\frac{K}{r}$ (according to (2)). If $\theta$ is defined as the angle of $\hat{\boldsymbol{\theta}}$ with $\boldsymbol{e}$, one gets:

$$\frac{K^2}{\alpha r} = 1 + e\cos\theta \qquad (e = |\boldsymbol{e}|). \qquad (7)$$

---

[8] In a short letter to Bernoulli in 1710, the mathematician Jakob Herman [9] proposed an elegant solution for the Kepler's problem. Starting from a figure quite similar to Newton's one (or our figure 1) he noted dd$x$ the projection of $cC$ on the $x$-axis and wrote Newton's law in the form $-a\mathrm{dd}x = xC^2/r^3$, where $C = xdy - ydx$ is twice the area of $SBC$ (i.e $C = K$ and $a = K^2/\alpha$ in our notations). Then, keeping one factor $C$ constant, he deduced that $-a\mathrm{d}x = C\frac{y}{r}$, which is the $x$-component of relation (5), when $\boldsymbol{e}$ is chosen along the $y$-axis (or when the $x$-axis is an axis of the conic as Herman supposed). In the same sentence, Herman also noted that $-a\frac{\mathrm{d}x}{x^2} = \frac{y}{x^2 r}(xdy - ydx)$ can be integrated as $\frac{a}{x} + \text{Cste} = \frac{r}{x}$, which is another quick way, although less natural and more technical than ours, to get the equation (7) of conics.

[9] As shown by Hamilton [10] on the basis of a pure geometrical reasoning following Newton, this hodograph is a circle. Thomson and Tait, cited in [10] p 331-333, recovered it in a way analogous to Herman (by integrating the relation $\boldsymbol{a} = -\frac{\alpha}{r^2}\hat{\boldsymbol{r}}\frac{xdy - ydx}{C}$).

Finally, the description of the Keplerian motion has been entirely solved since one has an explicit formula for the trajectory and an explicit statement (Kepler's second law) for the way it is traveled. Of course formula (7) has still to be mathematically studied (with the particular cases $e=1$ for a parabola, $0 \leq e < 1$ for an ellipse, $e > 1$ for a hyperbola). On figure 1, the $y$-axis has been taken to be parallel to the vector $\boldsymbol{e}$. Then $\theta$ is also the angular coordinate of a point of the trajectory (i.e. the angle between the $x$-axis and the vector $\hat{\boldsymbol{r}}$).

**5. Kepler's third law.**

The discrete time approach also leads to the exact formula for the orbital period. We proceed as usually in textbooks. $\mathcal{A}$ being the area of the ellipse, this period is simply

$$T = \frac{2\mathcal{A}}{K}, \qquad (8)$$

since $\frac{K}{2}$ is the area which is swept per unit time. If one admits that $\mathcal{A} = \pi ab$, where $a$ is in this section the semi-major axis given by

$$2a = r_{\min} + r_{\max} = r(\theta=0) + r(\theta=\pi) = \frac{K^2}{\alpha}(1-e^2)^{-1}, \qquad (9)$$

and $b = a\sqrt{1-e^2}$ is the semi minor axis[10], the period is simply:

$$T = 2\pi \alpha^{-1/2} a^{3/2}. \qquad (10)$$

Of course this value is expressed in terms of our somewhat arbitrary unit of time. But since the dimension of the right side of (10) is the correct one ($[\alpha] = [Length]^3 [Time]^{-2}$), this formula does not depend of the choice for this unit and is valid in the limit of a continuous trajectory.

**6. Conclusion.**

To sum up the above presentation of the derivation of Kepler's law from the gravitational force, one can say that it rests on the observation of a unique historical figure and on a series of equalities (3) to (5) which follow on from each other very simply and naturally. The figure recalls the definitions of velocity and acceleration (the simplest one in a first approach to Mechanics). It illustrates a simple algorithmic procedure with $\Delta t = 1$ and exhibits the law of equal areas for central forces, which are already important contributions of Newton. Our demonstration of the equation of the trajectory calls, neither for high brow geometry like Newton's one, nor for an expertise in vectorial differential equations (although (3) to (5) is equivalent to writing $\boldsymbol{a} = \frac{d\boldsymbol{v}}{dt} = -\frac{\alpha}{r^2}\hat{\boldsymbol{r}} = \frac{\alpha}{K}\frac{d\hat{\boldsymbol{\theta}}}{dt}$ as used for instance in [5,8]). Therefore this presentation, which continues the discrete time approach of Newton, is both an elementary and an efficient one. In particular, it can be addressed, either to students as an introduction to the concepts and methods of Mechanics, or to a large public interested by exact demonstrations concerning important historical issues.

---

[10] From equation (7) written as $r = p - ex \ (p = K^2\alpha^{-1})$ follows $x^2(1-e^2) + y^2 + 2epx - p^2 = 0$, or $X^2(1-e^2) + y^2 = \text{cste}$ by an adequate change of abscissa. This last relation immediately gives $b = a\sqrt{1-e^2}$ and $\mathcal{A} = \pi ab$ ($\pi a^2$ being the area of a circle of radius $a$).